\documentclass[twocolumn]{aastex62}
\usepackage{datetime}
\usepackage{appendix}
\usepackage{xcolor}
\def\msun{\rm M_\odot/h}
\def\mpc{\rm Mpc/h}
\def\zpeak{z_{\rm peak}}
\def\mpeak{M_{\rm peak}}

\def\anf{a_{\rm nf}}
\def\zf{z_{\rm f}}

\newdateformat{specialdate}{\THEYEAR-\twodigit{\THEMONTH}-\twodigit{\THEDAY}}

\begin{document}
\title{The Formation History of Subhalos and the Evolution of Satellite Galaxies}

\author{Jingjing Shi}
\affiliation{Kavli Institute for Astronomy and Astrophysics, Peking University, Beijing 100871, China}
\affiliation{Kavli IPMU (WPI), UTIAS, The University of Tokyo, Kashiwa, Chiba 277-8583, Japan}

\author{Huiyuan Wang}
\affiliation{Key Laboratory for Research in Galaxies and Cosmology, Department of Astronomy, University of Science and Technology of China, Hefei, Anhui 230026, China}
\affiliation{School of Astronomy and Space Science, University of Science and Technology of China, Hefei 230026, China}

\author{Houjun Mo}
\affiliation{Department of Astronomy, University of Massachusetts, Amherst MA 01003-9305, USA}
\affiliation{Astronomy Department and Center for Astrophysics, Tsinghua University, Beijing 10084, China}

\author{Mark Vogelsberger}
\affiliation{Kavli Institute for Astrophysics and Space Research, Massachusetts Institute of Technology, Cambridge, MA 02139, USA}

\author{Luis C. Ho}
\affiliation{Kavli Institute for Astronomy and Astrophysics, Peking University, Beijing 100871, China}
\affiliation{Department of Astronomy, School of Physics, Peking University, Beijing 100871, China}

\author{Min Du}
\affiliation{Kavli Institute for Astronomy and Astrophysics, Peking University, Beijing 100871, China}

\author{Dylan Nelson}
\affiliation{Max-Planck-Institut f\"{u}r Astrophysik, Karl-Schwarzschild-Str. 1, 85741 Garching, Germany}

\author{Annalisa Pillepich}
\affiliation{Max-Planck-Institut f\"{u}r Astronomie, K\"{o}nigstuhl 17, 69117 Heidelberg, Germany}

\author{Lars Hernquist}
\affiliation{Harvard-Smithsonian Center for Astrophysics, 60 Garden Street, Cambridge, MA 02138}

\correspondingauthor{Jingjing Shi \& Huiyuan Wang}
\email{jingssrs1989@gmail.com; whywang@ustc.edu.cn}

\begin{abstract}
Satellites constitute an important fraction of the overall galaxy population and are believed to form in dark matter subhalos. Here we use the cosmological hydrodynamic simulation TNG100 to investigate how the formation histories of subhalos affect the properties and evolution of their host galaxies. We use a scaled formation time ($\anf$) to characterize the mass assembly histories of the subhalos before they are accreted by massive host halos.
We find that satellite galaxies in young subhalos (low $\anf$) are less massive and more gas rich, and have stronger star formation and a higher fraction of ex situ stellar mass than satellites in old subhalos (high $\anf$). Furthermore, these low $\anf$ satellites require longer timescales to be quenched as a population than the high $\anf$ counterparts. We find very different merger histories between satellites in fast accretion (FA, $\anf<1.3$) and slow accretion (SA, $\anf>1.3$) subhalos. For FA satellites, the galaxy merger frequency dramatically increases just after accretion, which enhances the star formation at accretion. While, for SA satellites, the mergers occur smoothly and continuously across the accretion time. Moreover, mergers with FA satellites happen mainly after accretion, while a contrary trend is found for SA satellites. Our results provide insight into the evolution and star formation quenching of the satellite population.
\end{abstract}


\section{Introduction}
In the prevailing paradigm of galaxy formation, galaxies grow in dark matter halos, built via gravitational collapse of small density peaks in the early Universe. Within halos, baryonic gas cools radiatively and condenses. The gas is then converted into stars, forming galaxies \citep{ReesOstriker1977,WhiteRees1978,FallEfstathiou1980}. Halos grow hierarchically by accreting smaller structures. These small structures become substructures (subhalos) of host halos, and galaxies within them become satellite galaxies, orbiting around the central galaxies residing in the deep gravitational potential wells of the host halos. Many studies use various methods to determine the satellite abundance and suggest that about 30\% of galaxies with stellar mass $\sim 10^9 \msun$ are satellites, while the satellite fractions decrease with increasing stellar mass \citep{Mandelbaum2006,Tinker2007,VandenBosch2008,Yang2008,Wetzel2013,Bray2016}. Satellite galaxies apparently constitute a non-negligible population of the entire galaxy population. 

After being accreted, satellites are believed to experience some specific processes that can quench ongoing star-formation or even cause morphological transformation. For example, subhalos and satellite galaxies stop growing, or lose their mass due to the strong tidal stripping of host halos \citep[e.g.][]{GunnGott1972}; some satellites even get tidally disrupted (e.g. into intra-cluster light) before merging with the centrals \citep{Mihos2005,Wetzel2010,Bahe2019}; dynamical friction gradually reduces the orbital angular momentum so that the satellites sink towards the halo center \citep{Chandrasekhar1943a,Chandrasekhar1943b,Chandrasekhar1943c}; ram pressure can remove the hot halo gas reservoir or even the cold gas in the disk of satellites \citep{GunnGott1972,Ayromlou2019,Yun2019}; `harassment' from neighbouring galaxies can tidally heat the satellite system \citep{FaroukiShapiro1981}; satellite-satellite mergers can affect both the star formation activity and morphology \citep{MakinoHut1997,Wetzel2009}. However, the importance of these processes in quenching/transforming satellites, and their dependence on various halo properties is still under debate.  

In observations, satellites have been studied in great detail. \cite{VandenBosch2008} investigated the efficiency of satellite specific processes by comparing the color and concentration of satellites and centrals of the same stellar mass at $z=0$ (see also \citealt{Peng2010,Peng2012,Wang2018}). These studies assume that centrals at $z=0$ closely resemble the progenitors of satellites at the time when they are being accreted by host halos. However, such assumptions ignore the evolution of galaxies (see for example \citealt{Pannella2009,Newman2012,Stark2013,Barro2017,Genzel2017,Tadaki2017,Price2019}). Subsequently, \cite{Yang2012} and 
\cite{Wetzel2013} took evolution into account and used an empirical parameterization method to derive the initial star formation rate (SFR) of satellites at the accretion time, and further investigated the quenching time scale of satellites.
In their methods, it is assumed that at the time of accretion, satellite progenitors resemble central galaxies of the same stellar mass. This assumption, however, still needs to be verified. 

These satellite specific processes can be investigated in theoretical galaxy formation models, such as semi-analytical models (SAMs), hydrodynamical simulations, and halo-based models. In SAMs, these processes are included in a parameterized way. The efficiencies and scalings that characterize these processes are constrained by fitting simultaneously the observed stellar mass functions and quenched fractions at various redshifts (see \citealt{Henriques2015} for example). These models reasonably reproduce the observed red fractions of galaxies as a function of stellar mass in varying environments across several redshifts, the satellite passive fractions as a function of halo mass, and the projected distance from the central galaxies, and the clustering signal for blue and red galaxies \citep{Henriques2017}. However, significant differences between model predictions and observations are also presented in the literature \citep{Hirschmann2014,WangEnci2018a,WangEnci2018b}. These studies suggest that the SAMs require an improvement in the environmental processes in order to reduce the differences between centrals and satellites. In fact, the detailed treatment of the environmental processes varies from model to model. For example, the earlier versions usually assume an instantaneous stripping of hot gas around satellites once they fall into their hosts (see \citealt{Croton2006,Bower2006}), while subsequent models allow for a gradual loss of the hot gas \citep{Font2008,Kang2008,Weinmann2010,Guo2011,Hirschmann2014}.

Halo-based models have often been used to study the connection 
between galaxies and dark matter halos 
\citep[e.g.][]{Jing1998,Berlind2002,Yang2003}.
In the subhalo abundance matching methods (SHAMs, see \citealt{Kravtsov2004,ValeOstriker2004,Conroy2006}), 
galaxies are populated within subhalos and a tight correlation between galaxy stellar mass and halo properties, such as halo mass, is adopted. However, the halo mass-based SHAMs fail to reproduce small scale clustering measurements, which are usually dominated by satellites 
\citep{Yang2012,Behroozi2013,Moster2013,Rodriguez-Puebla2017}.
\cite{Campbell2018} suggested that the discrepancy could be caused by some simple assumptions that are not valid in reality. For example, the models neglect the contribution of `orphan' galaxies, the mass growth of satellites after their accretion, and the influence of halo assembly history. In fact, persistent star formation after accretion is found to be important for understanding the bimodality of satellites \citep{Weinmann2009,Kang2008,WangYu2007,Simha2009,Wetzel2009}. 
In addition, tidal stripping of stars \citep{Mihos2005} and satellite-satellite mergers \citep{Wetzel2009} may be also important since they alter the SMHM relations.

All satellite-specific processes, such as tidal stripping, ram pressure, and galaxy interactions, are environmental effects that can be simulated by solving the equations for collisionless dynamics and hydrodynamics (see \citealt{Vogelsberger2019} for a review on cosmological simulations). In particular, hydrodynamical simulations can account for these satellite-specific processes in a self-consistent manner, although the results should be interpreted with caution because of limited numerical resolutions and uncertainties in subgrid models. The predicted satellite stellar mass function at a given halo mass from simulations matches local observations \citep{Bahe2017}. \cite{Wright2019} studied the quenching timescales of galaxies in EAGLE simulations for centrals and satellites, finding that different physical mechanisms are at play for galaxies of low, intermediate, and high mass. \cite{Bahe2015} showed that ram pressure stripping is the main mechanism that is responsible for satellite quenching in groups and clusters using GIMIC cosmological hydrodynamical simulations. \cite{Yun2019} and \cite{Jung2018} studied the efficiency of ram pressure in removing the gas content of satellites using TNG100 and HorizonAGN. Joshi et al. (in preparation) demonstrates with the TNG50 and TNG100 simulations that group and cluster environments transform galaxies from disc to spheroids, possibly because of tidal shocking. Using different suites of hydrodynamical simulations, others have shown how the disruption rate, tidal mass loss, specific star formation rate, luminosity-weighted age, stellar metallicity, and alpha element abundance ratios of satellites depend on the time since infall \citep{Rhee2017,Pasquali2019,Bahe2019}. \cite{Engler2020} show that satellite and central galaxies lie on distinct average relations between stellar mass and current dynamical mass in TNG100.
However, most previous works focus on how satellites evolve after accretion, without considering in detail the role of the formation history of the subhalos prior to accretion.

More recently, \cite{Shi2018} studied the formation history of subhalos before they are accreted by more massive host halos. Interestingly, they find that there are two infall subhalo populations, which correspond to the fast-accretion and slow-accretion phases found in normal distinct halos \citep{Zhao2003}. They also compared these infall subhalos with normal halos at the time of accretion and found that infall subhalos are usually younger than normal halos of the same halo mass. The different halo growth histories of those two infall subhalo populations may leave an imprint on the satellites at $z=0$. In addition, the difference between infall subhalos and normal halos raises the question of whether the satellite progenitors before accretion are different from general centrals at the time of accretion. In this work, we use the cosmological hydrodynamical simulations IllustrisTNG to carry out investigations on satellite galaxies and infall subhalos, focusing on how the formation history of infall subhalos influence the initial status of satellites and their further evolution.

The structure of the paper is organized as follows. In section \ref{sec_method}, we briefly introduce the IllustrisTNG simulations, our sample selection, and definitions adopted in this work. In section \ref{sec_res}, we present our results, focusing on the role of the formation history on satellites/the difference between satellites and centrals at the time of accretion, evolution after accretion, and the state at z=0. In section \ref{sec_dis} we discuss the implications of our work for SHAMs and satellite quenching. We summarize our main results in section \ref{sec_sum}.

\section{Method}
\label{sec_method}

\subsection{The {\rm ILLUSTRIS-TNG} Simulations}
\label{subsec_sim}

Throughout this work, we use the data from the IllustrisTNG simulations \citep{Naiman2018,Marinacci2018,Springel2018, Pillepich2018, Nelson2018, Nelson2019a, Pillepich2019} that feature a novel model for AGN feedback, magneto-hydrodynamics, a new scheme for galactic winds, and updated choices for stellar evolution and chemical enrichment (for more details on the TNG model, see  \citealt{Weinberger2018, Pillepich2018, Nelson2019b}), leading to significant improvements over the original Illustris model \citep{Vogelsberger2013,Vogelsberger2014,Vogelsberger2014nature, Genel2014}. The simulations are calibrated to reproduce well the cosmic star formation rate density at $z\leq10$, the observed galaxy stellar mass function, the stellar-to-halo mass relation, the total gas mass content within the massive groups, the stellar mass - stellar size relation, and the BH - galaxy mass relations at $z=0$ \citep{Pillepich2018}. Besides those used to calibrate the models, the simulations can reproduce well the galaxy stellar mass functions up to $z\sim 4$ \citep{Pillepich2018smhmr}, the galaxy clustering of blue and red galaxies \citep{Springel2018}, stellar sizes up to $z\sim2$ \citep{Genel2018}, etc..
To study the evolution of satellite galaxies, we use the highest resolution run of the $\sim 100$Mpc box TNG100-1 (TNG100 hereafter), which is performed with the moving-mesh code AREPO \citep{Springel2010,Weinberger2019} in a periodic box of $75\mpc$ on a side. It follows the dynamical evolution of ${\rm 1820}^3$ DM particles and approximately ${\rm 1820}^3$ gas cells or stellar/wind particles from $z=127$ to $z=0$, producing $100$ snapshots between $z=20$ and $z=0$. The DM particle mass is $5.1\times 10^6\msun$ and the average gas cell mass is $9.4\times 10^5\msun$. The cosmological parameters of the IllustrisTNG simulations are consistent with recent Planck measurements \citep{Planck2016}: $\Omega_{\rm m}=0.3089$, $\Omega_{\rm b}=0.0486$, $\Omega_{\rm \Lambda}=0.6911$, $\sigma_8=0.8159$, $n_s=0.9667$ and $h=0.6774$. 

Halos and structures within them, for example galaxies, are identified by using the Friends-of-Friends (FOF) and SUBFIND algorithms \citep{Davis1985,Springel2001}. Usually, each FOF group contains one or more SUBFIND structures (hereafter substructure or subhalo) and the baryonic component in a substructure is defined as a galaxy. In this paper, FOF halos are referred to as halos or host halos. The most massive substructure in a FOF halo is classified as a central subhalo and its galaxy is regarded as the central galaxy of the halo.  The other substructures (if they exist) are called satellite subhalos, the baryonic components of which are referred to as satellite galaxies. We verified that our results do not change when we select satellites that are within the virial radius of host halos. To be concise, we use the term ``satellites (or centrals)" to refer to both satellite (or central) galaxies and satellite (or central) subhalos in the following text.

The subhalos (including both central and satellite subhalos) are connected across the 100 snapshot outputs by the SUBLINK algorithm \citep{Rodriguez-Gomez2015}. The descendant of a subhalo is identified by matching the weighted baryons in the next snapshot (i.e. lower redshift), where the most bound particles/cells have the highest priorities. Each subhalo has only one descendant, yet it can have more than one progenitor. The main progenitor is defined as the one with the `most massive history' behind it \citep{DeLucia2007}. In this way, merger trees are constructed. 

We also use the data from the original Illustris simulation and TNG300-1 to verify the dependencies of our results on the parameterized sub-grid physics and the numerical resolution. Most of our results remain the same across the three sets of simulations.

\subsection{Sample and galaxy properties}
\label{subsec_sample}

We first describe the definitions of some common parameters used in this paper for halos and galaxies.
\begin{itemize}
    \item \textit{Halo mass:} the mass contained in the spherical region (centered on the most bounded particle) where the mean mass density is equal to $200\rho_{\rm crit}(z)$. Note that halo mass is only measured for central subhalos. 
    \item \textit{Galaxy stellar mass:} the sum of all stellar particles within twice the stellar half mass radius, $2R_\star$, where $R_\star$ is defined using all stellar particles within the subhalo.
    \item \textit{Galaxy gas mass:} the sum of all the mass in gas cells within $2R_\star$.
    \item \textit{Specific star formation rate (sSFR):} star formation rate (SFR) per unit stellar mass, where SFR is computed by summing up the star formation rates in all star-forming gas cells within $2R_\star$.
    \item \textit{Ex situ stellar mass fraction $f_{\rm ex\ situ}$:} the fraction of stellar mass that is formed in other galaxies and accreted later by the galaxy \citep{Rodriguez-Gomez2017}. This parameter can be used to quantify the contribution of galaxy mergers to the mass growth.
\end{itemize}

\begin{figure}
\centering
 \includegraphics[width=0.9\linewidth]{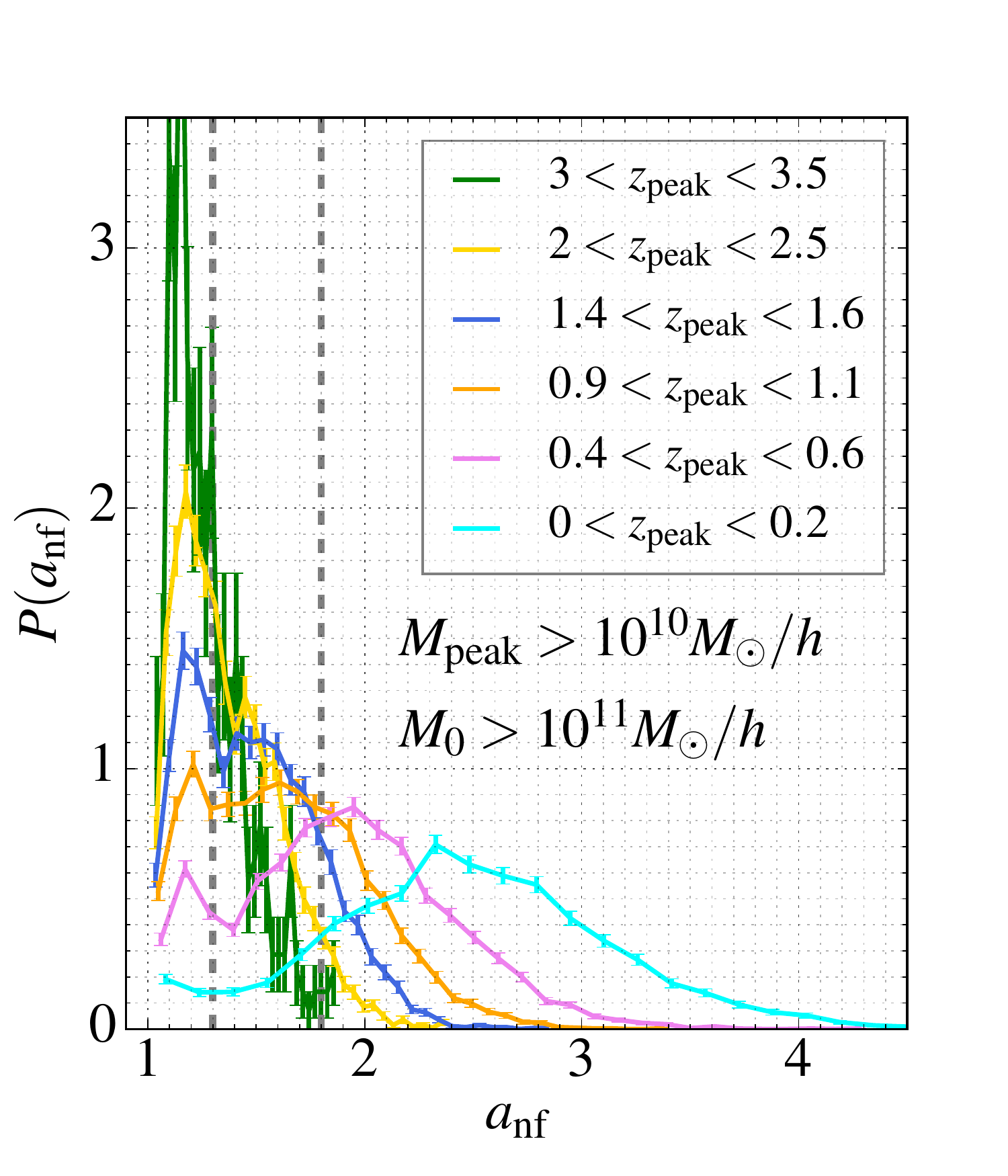}
\caption{\label{fig_bimodal} Probability distribution of the scaled formation time, $\anf\equiv(1+\zf)/(1+\zpeak)$, for TNG100 satellite subhalos that survived at $z=0$ and are accreted at different redshifts, where $\zf$ is the formation time and $\zpeak$ is the accretion time, as defined in section \ref{subsec_sample}. Here we include only the satellites with $\mpeak>10^{10}\msun$ and host halo mass $M_{0}>10^{11}\msun$. The gray dashed vertical lines indicate the division thresholds for the FA, SA1, and SA2 populations. The error bars are Poisson errors.}
\end{figure}

We select all satellite galaxies at $z=0$ and trace them back in time along their merger trees. Below, we list some parameters and phrases, which can record the evolution history of satellite galaxies and subhalos, and describe them briefly:
\begin{itemize}
    \item \textit{Accretion time $\zpeak$:} the redshift when the subhalo of a satellite galaxy is a central subhalo and its halo mass reaches the maximum during its lifetime. Therefore, $\zpeak$ is usually thought to be the time when the subhalo starts to be influenced by the gravitational field of its host halo.
    \item \textit{Peak halo mass $\mpeak$:} the halo mass at $\zpeak$.
    \item \textit{Satellite and Central phases:} The central phase is the time when a $z=0$ satellite galaxy is a central galaxy (i.e. at $z>\zpeak$), and the satellite phase is the time after being accreted by a more massive host halo (i.e. at $z<\zpeak$).
    \item \textit{Formation time $\zf$ and $\anf$:} $\zf$ is the redshift when the subhalo of a satellite galaxy is a central subhalo and its subhalo reaches half of its peak mass, $\mpeak/2$, for the first time. Also, $\anf\equiv (1+\zf)/(1+\zpeak)$ is the formation redshift scaled by the accretion redshift. $\zf$ and $\anf$ therefore reflect the formation history of a subhalo during its central phase.
    \item \textit{FA, SA1 and SA2 populations:} Subhalos with $a_{\rm nf}<1.3$ are classified as the fast accretion (FA) population; those with $a_{\rm nf}>1.3$ are classified as the slow accretion (SA) population \citep{Shi2018}. As we do not know whether they grow because of accretion (e.g. of ``smooth" DM) or assembly (i.e. of merging with other DM haloes), we could stick to one name but then it could mean both physical processes. The SA population is further divided into SA1 ($1.3<a_{\rm nf}<1.8$ ) and SA2 ($a_{\rm nf}>1.8$) samples. The FA population corresponds to the first peak in the bimodal distribution of $\anf$ (see Figure \ref{fig_bimodal}), while the SA population corresponds to the second peak in $\anf$ distribution. We adopt the transition values from the first to the second peaks, $\anf\sim 1.3$, which are independent of $\zpeak$, as a division of the two populations. The further division of the SA populations using $\anf=1.8$ makes the three subsamples to have comparable numbers of galaxies. By comparing the FA population with the SA population, and the SA1 population with SA2 population, we can figure out whether there exists a radical difference between the FA and SA populations, or the difference is purely an influence of the increasing $\anf$. See Section \ref{subsec_bimodal} for more details. 
    
\end{itemize}

The quantities with subscript `0' denote the values measured at $z=0$ and those with  subscript `peak' denote the values at $\zpeak$.
For example, $M_0$ and $\mpeak$ denote the halo mass at $z=0$ and $\zpeak$ respectively, while $M_{\rm \star,0}$ and $M_{\rm \star,peak}$ mean the stellar mass at $z=0$ and $\zpeak$ respectively. In this paper, we mainly focus on satellite galaxies residing in host halos of $M_{0}>10^{11}\msun$ at $z=0$ and having peak halo mass $\mpeak>10^{10}\msun$.  We note that a galaxy that is identified as a satellite at $z=0$ was actually a central galaxy in its central phase. \textit{To avoid any confusion, we always call it a satellite galaxy, even in its central phase.}
In this paper, we will compare satellites with centrals at both $z=0$ and $\zpeak$. The central galaxies at $z=0$($\zpeak$) shown in the following figures are galaxies classified as centrals at $z=0$($\zpeak$),  regardless of whether or not they are classified as centrals at other redshifts.

\section{Results}
\label{sec_res}

The main purpose of this paper is to investigate whether and how, for a satellite at $z=0$, its formation history in its central phase affects or leaves an imprint on its evolution in its satellite phase and final state, and to understand the different evolutionary paths between centrals and satellites. In Section \ref{subsec_bimodal}, we first show the formation time distribution for these satellites and the separation of fast and slow accretion populations. In Section \ref{subsec_smhmr}, we present results on the stellar mass - halo mass (SMHM) relation. The evolution of stellar mass, gas content, and star formation are presented in Section \ref{subsec_stellar_assembly} and Section \ref{subsec_gas_sf}. We also present the satellite quenched fraction in \ref{subsec_fq}. Finally, in Section \ref{subsec_gal_merger}, we present results about galaxy mergers for these satellite galaxies.

\subsection{Bimodal formation time distribution of infall subhalos}
\label{subsec_bimodal}

In our previous work \citep{Shi2018}, we found that the probability distribution of the scaled formation time ($\anf$) for the satellite subhalos is bimodal at a given accretion time $\zpeak$, using a cosmological N-body simulation. The result is confirmed here with the cosmological hydrodynamical simulation TNG100, as shown in Figure \ref{fig_bimodal}. Our tests in \cite{Shi2018} with much larger N-body simulations suggest that such bimodality is independent of the host halo mass at $z=0$ and depends only weakly on peak halo mass $\mpeak$. 

In \cite{Shi2018}, we demonstrated that the bimodality in normalized formation time $\anf$ is closely connected to the two accretion phases (fast and slow accretion phases) found in the mass accretion history of dark matter halos \citep{Zhao2003}.  The peak with small $\anf$ corresponds to the subhalos that are in the fast-accretion phase (i.e. younger), while the population with large $\anf$ corresponds to the subhalos that are in the slow-accretion phase (i.e. older). It should be noted that the infall time distribution of satellite galaxies selected to be found within the virial radius from a more massive host is bimodal as well, even when infall is defined as virial radius crossing. In that case, the bimodality is removed when backsplash satellites are included, i.e those are outside the virial radius at given time (\citealt{Yun2019,Engler2020}.
Since the $\anf$ distribution of FA halos is narrow and its peak is always around $1.2$, almost independent of $\zpeak$, host halo mass, and $\mpeak$ \citep{Shi2018}, we adopt the demarcation value $\anf=1.3$ to divide them into FA and SA populations (see Figure \ref{fig_bimodal}). The fractions of the two populations are a strong function of $\zpeak$. Satellite subhalos that are accreted at $\zpeak>2.3$ (roughly $11$ Gyr ago) are mainly FA halos, while those accreted later are dominated by SA halos.

As we will show below, satellites in the FA and SA satellite galaxies evolve in different ways, in particular in the satellite phase. There are two possible origins for the difference. One is that FA and SA satellites are distinct populations, as suggested by the bimodal $\anf$ distribution. The other is that the difference simply reflects a continuous change as a function of $\anf$, since they, by definition, have different formation times. To distinguish the two possibilities, in principle, one should divide the sample into many small $\anf$ bins, then check how the satellite properties and their evolution vary with $\anf$. However, due to the limited number of satellites (as shown in Table \ref{tabel_1}), we adopt a coarse binning method, in which the FA population is kept unchanged and the SA population is split into two subsamples, SA1 ($1.3<a_{\rm nf}<1.8$) and SA2 ($a_{\rm nf}>1.8$).
We choose this approach because the FA population has a much narrower $\anf$ distribution than the SA population. The demarcation value of $1.8$ is chosen so that the three populations have roughly a comparable number of galaxies (see the other vertical dashed line in figure \ref{fig_bimodal}). Table \ref{tabel_1} briefly lists the number of satellites in each subsample. If a satellite property changes smoothly across the three samples, i.e. FA, SA1, and SA2, the property is very likely determined by halo formation history. If a special feature for a property appears in FA, but is totally absent in SA1 and SA2, it implies that FA and SA are indeed different in this property.

\begin{table}[h!]
\centering
 \caption{Number of satellites in each subsample of varying $\zpeak$ bins that correspond to different lines in Figure \ref{fig_bimodal}.}
 \label{tabel_1}
\begin{tabular}{|c|c|c|c|} 
 \hline
 $\zpeak$ & FA & SA1 & SA2 \\ [0.5ex] 
 \hline\hline
 $0-0.2$ & 186 & 367 & 3154 \\ 
 \hline
 $0.4-0.6$ & 754 & 1624 & 3091 \\
 \hline
 $0.9-1.1$ & 1181 & 2277 & 1627 \\
 \hline
 $1.4-1.6$ & 1518 & 2285 & 720 \\
 \hline
 $2-2.5$ &  2021 & 2097 & 228 \\ [1ex] 
 \hline
 $3-3.5$ & 329 & 165 & 5 \\ [1ex] 
 \hline
\end{tabular}
\end{table}

\subsection{Stellar mass - halo mass relation}
\label{subsec_smhmr}

\begin{figure*}
\centering
 \includegraphics[width=1.\linewidth]{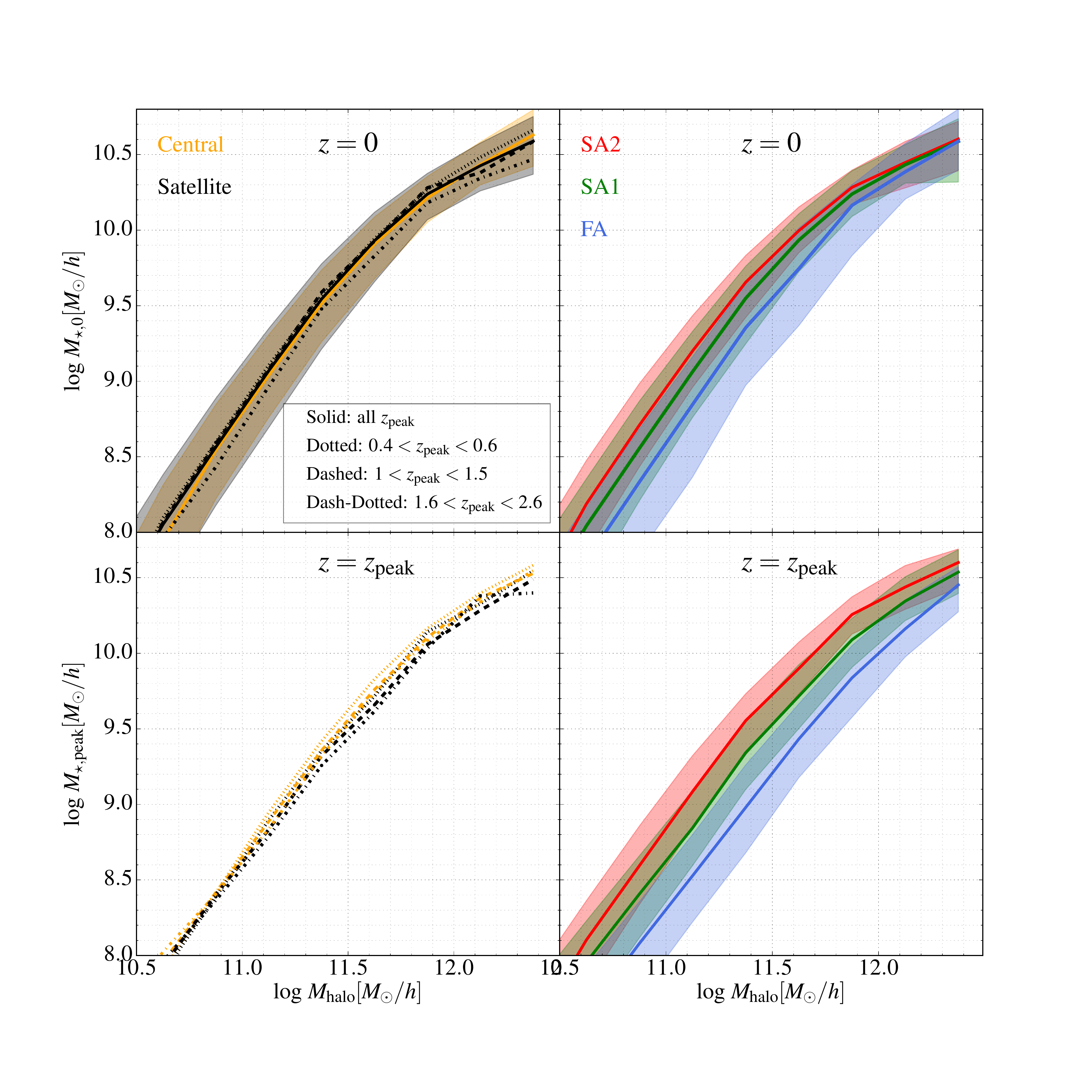}
\caption{\label{fig_smhm} Stellar mass to halo mass relation from TNG100. Here $M_{\rm halo}$ is $M_{0}$ for centrals and $\mpeak$ for satellites. 
Upper left: the $z=0$ SMHM relation for the satellites (black lines) and centrals (orange lines). The dotted, dashed, and dash-dotted lines show the relation for satellites accreted at different $z_{\rm peak}$ bins, as indicated by the label.  
Lower left: The $\zpeak$ SMHM relation for satellites at three $\zpeak$ bins and centrals at the corresponding redshifts. See the text for more details.
Upper right: the median $z=0$ SMHM relation for FA (blue), SA1 (green), and SA2 (red) satellites. The shaded area shows the $1 \sigma$ dispersion around the median relations. 
Lower right: The SMHM relation at $\zpeak$ for satellites. 
Color and line types are the same as the upper right panel.}
\end{figure*}

The stellar mass-halo mass (SMHM) relation can provide valuable constraints on the efficiency in transforming gas into stars and has been studied in great detail in the literature (e.g. \citealt{Yang2003,Zheng2007, Yang2008, Yang2009, Behroozi2010,Guo2010,WangJing2010,Yang2012,WechslerTinker2018,Pillepich2018smhmr}). In this subsection, we present the SMHM relations for the three samples of satellites and central galaxies. In order to understand the evolution of the relations, two kinds of SMHM relations are investigated as shown in Figure \ref{fig_smhm}. The first one is the $z=0$ SMHM relation, in which halo mass ($M_{\rm halo}$) is taken as the halo mass at $\zpeak$ ($\mpeak$) for satellites and halo mass at $z=0$ ($M_{\rm 0}$) for centrals.
The second one is the $\zpeak$ SMHM relation. In this re- lation, for both centrals and satellites, stellar mass and halo mass at zpeak are adopted.

We first show the $z=0$ SMHM relations for satellite and central galaxies in the upper-left panel of Figure \ref{fig_smhm}. To check the dependence on accretion time, we select three $\zpeak$ ranges ($0.4<\zpeak<0.6$, $1.0<\zpeak<1.5$ and $1.6<\zpeak<2.6$), and show results in these $\zpeak$ ranges in the same panel.
Since the simulation box is quite small, we only show the SMHM relation in a relatively narrow halo mass range (from about $10^{10.5}$ to $10^{12.5}\msun$). As one can see, for both centrals and satellites, $M_{\star,0}$ is strongly correlated with halo mass, with a scatter of about $0.25$ dex. It is in good agreement with previous studies \citep{Yang2009,More2011,Moster2013,Behroozi2013,ZuMandelbaum2015,Matthee2017}. Moreover, the SMHM relations for satellites are consistent with that for centrals, and almost independent of the accretion time $\zpeak$, as shown by the non-solid lines. 
We then show the $\zpeak$ SMHM relations at three $\zpeak$ ranges in the lower-left panel. Similar to the $z=0$ relation, centrals and satellites almost follow the same relation. Furthermore, there is only weak evolution with redshift, which is consistent with the independence of the $z=0$ SMHM relation on accretion time. All of these results suggest that host halo mass (or $\mpeak$) is the major factor that governs the galaxy stellar mass over a wide range of redshift for both centrals and 
satellites. Note that all this applies when the halo mass of satellites are considered at $\zpeak$, i.e. before any environmental effect. In fact, the $z=0$ relations between current stellar and dynamical mass are clearly distinct for satellite and central galaxies \citep{Engler2020}.

We then investigate the $\zpeak$ SMHM relation for the three satellite samples with different $\anf$ separately  (the lower right panel of Figure \ref{fig_smhm}). 
At a given halo mass, $M_{\rm \star,peak}$ increases gradually from the lowest $\anf$ sample (FA, youngest) to the highest $\anf$ sample (SA2, oldest). This suggests that the $\zpeak$ SMHM relation changes continuously with $a_{\rm nf}$. In addition, the $a_{\rm nf}$ dependence slightly decreases with increasing $\mpeak$. 
Our further tests show that the $\zpeak$ SMHM relations for the three samples are all independent of $\zpeak$, consistent with the results shown in the lower left panel of Figure \ref{fig_smhm}. Recently, several works \citep{Matthee2017,Zehavi2018,Artale2018} studied the output of SAMs and hydrodynamical simulations and found that, at fixed halo mass, old halos tend to host more massive central galaxies at $z=0$. For instance,  \cite{Artale2018} found that the oldest $20$\% halos host galaxies that are $\sim 0.4$ dex larger than the youngest $20$\% halos. This is broadly consistent with our findings. 

At $z=0$, the difference among the three satellite samples is apparently reduced, but still significant (upper right panel of Figure \ref{fig_smhm}). The maximum difference of about $0.25$ dex appears at $\mpeak<10^{11.5}\msun$ and the difference decreases with increasing $\mpeak$.
At $\log\mpeak\sim 11.9$ ($\log M_{\rm \star,0}\sim 10.2$), the difference becomes negligible. 
We also check the dependence of the $z=0$ SMHM relation on accretion time and no significant dependence is found. By comparing the $\zpeak$ and $z=0$ SMHM relations, we find that satellites grow with time in the satellite phase, but the growth amplitude decreases with increasing $\anf$.  
    We will discuss this in more detail in the following. These results suggest that there is a time delay between halo assembly and galaxy assembly. So even when a halo becomes a substructure of another massive halo and starts to lose its mass, the galaxies within it can still grow (see also Fig. 8 of \citealt{Engler2020}).

\begin{figure*}
\centering
\includegraphics[width=0.9\linewidth]{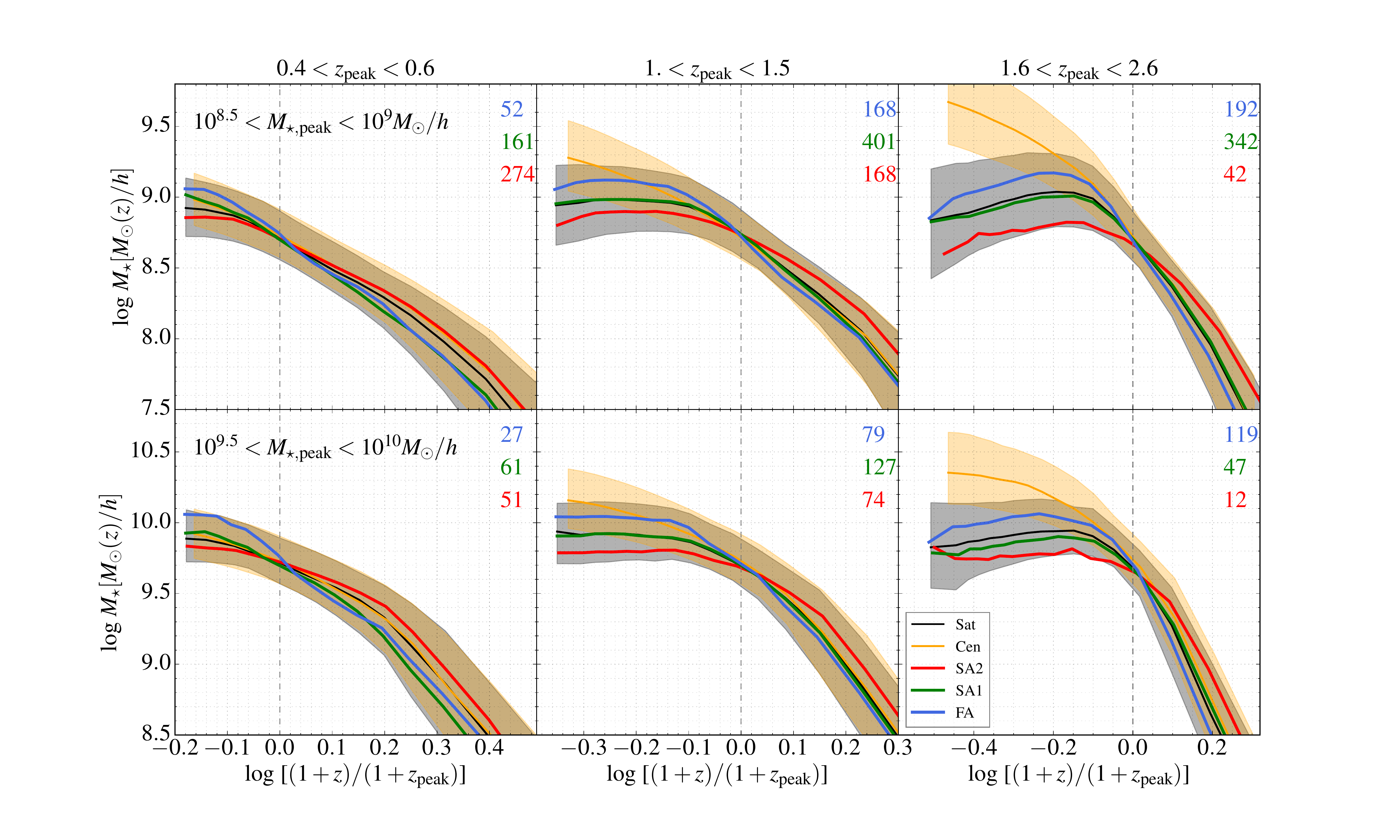}
\caption{\label{fig_mstarz} Stellar mass assembly of TNG100 galaxies. The solid lines show the median stellar mass evolution for different satellite samples as indicated in the panels and the corresponding centrals. The vertical gray dashed lines indicate the separation of the satellite and central phase for the infall satellite galaxies.
These central galaxies have stellar masses similar to the satellites, and are classified as centrals at a redshift close to $\zpeak$, regardless of whether or not they are classified as centrals at other redshifts. The shaded black and orange areas show the 1$\sigma$ dispersion of the evolution for the satellites and centrals. The numbers in each panel show the number of satellites that we have for each subsample. The same statistics applies to Figure \ref{fig_sfrz}, \ref{fig_fqz}, and \ref{fig_faccz}.}
\end{figure*}

\subsection{Stellar mass assembly}
\label{subsec_stellar_assembly}

To see more clearly how the stellar mass of satellites evolves with time, we select satellites with two $M_{\rm \star,peak}$ ranges and three $\zpeak$ ranges and present their stellar mass evolution in Figure \ref{fig_mstarz}.
For comparison, we also show the evolution of the corresponding central galaxies in the same figure. These central galaxies have stellar masses comparable to these satellites, and are classified as centrals at redshifts close to $\zpeak$, regardless of whether or not they are classified as centrals at other redshifts. We first check the mass growth of these satellite galaxies in their central phase (i.e. $z>\zpeak$). The mass growth for satellites as a whole (black solid lines) resembles that for centrals (orange solid lines). Moreover, satellites with high $\anf$ (older) grow at a longer timescale and more slowly than centrals, while satellites with low $\anf$ (younger) grow faster than the corresponding centrals. 

At the beginning of the satellite phase, satellites continue to grow, although the growth rate appears to be slowed down.  On average, the satellites stop growing at $\log{[(1+z)/(1+\zpeak)]}\sim-0.1$, which corresponds to a time scale from $1.3$ Gyr (at $z=2$) to $2.7$ Gyr (at $z=0.5$). The halo dynamical time scales, defined as $R_{\rm 200crit}/V_{\rm crit200}=0.1/H(z)$, at $z=2$ and $z=0.5$ are about $0.5$ Gyr and $1.1$ Gyr, so the mass growth ceases roughly at $2$ times the dynamical time scale after accretion, likely due to the lack of gas supply (see Section \ref{subsec_gas_sf}). For satellites that are accreted at low redshift ($\zpeak\sim0.5$ and $\zpeak\sim1.2$), the stellar mass remains unchanged after reaching its peak; while the satellites accreted at high redshift (e.g. $\zpeak\sim2$) eventually lose mass by a factor of $0.2$ dex after reaching a maximum. This is possibly due to tidal stripping, since these satellites are accreted earlier and tend to reside in the inner region of their host halos at $z=0$, where the tidal field is the strongest \citep{Gnedin1999}. After being accreted, satellites with smaller $\anf$ (younger) apparently grow faster and more (about $0.4$ dex) than satellites with larger $\anf$ (older), ranging from $0.15$ to $0.25$ dex. The growth rate decreases with increasing $a_{\rm nf }$, consistent with the results shown in the SMHM relations (see the left panels of figure \ref{fig_smhm}). This is saying that the stellar mass growth relies on $\anf$ in both central and satellite phases, and the difference between FA and SA populations does not relate to the bimodality of $\anf$.

\subsection{Gas content and star formation}
\label{subsec_gas_sf}

\begin{figure*}
\centering
\includegraphics[width=1.\linewidth]{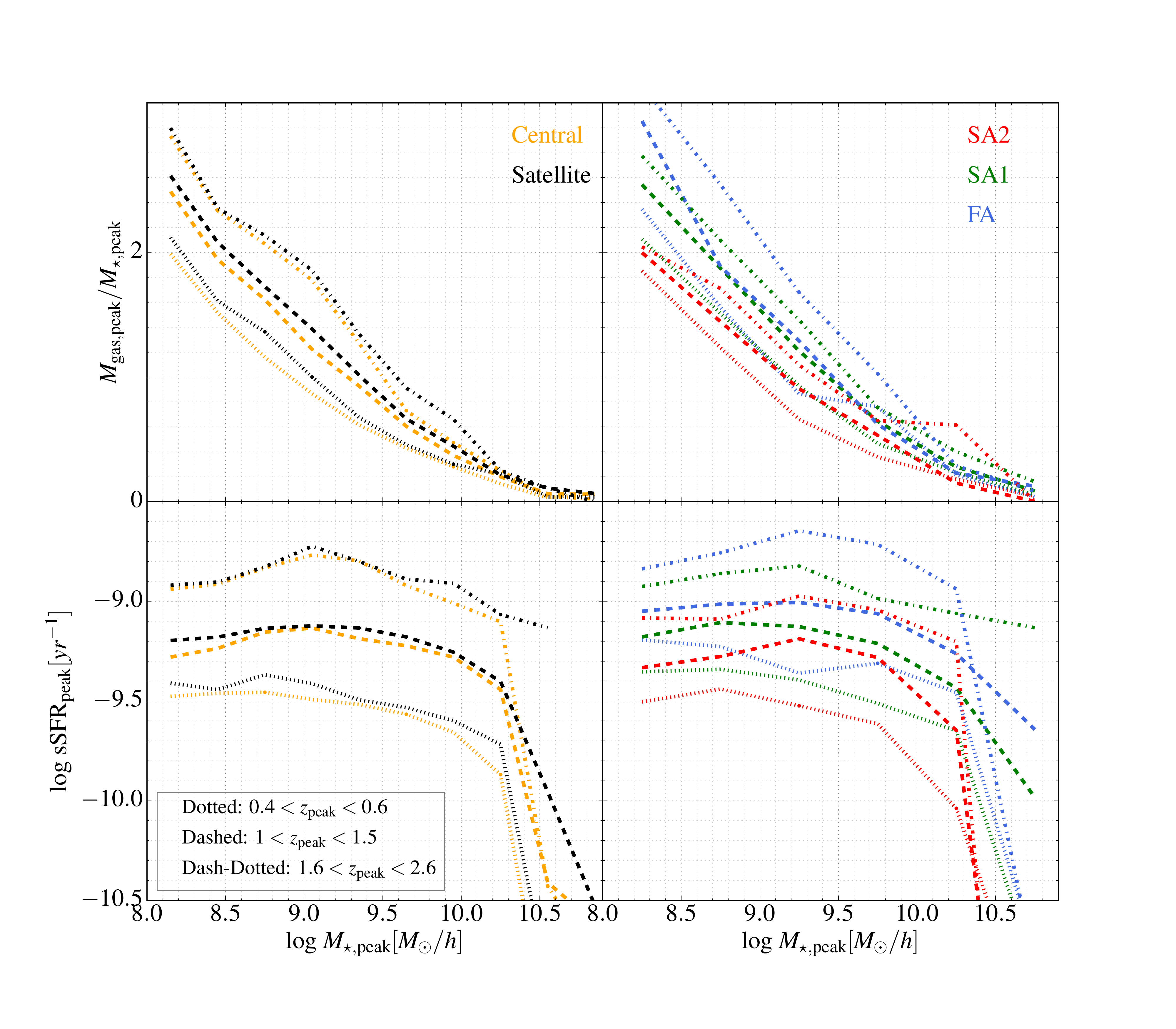}
\caption{\label{fig_prop_zpeak} Median gas mass fractions and sSFRs in TNG100 galaxies. Left column: gas to stellar mass ratio (top) and sSFR (bottom) for centrals (orange) and satellites (black) in various $\zpeak$ bins. 
Right column: gas to stellar mass ratio (top) and sSFR (bottom) as a function of the stellar mass at $z_{\rm peak}$ for the three satellite populations in various $\zpeak$ bins. }
\end{figure*}

\begin{figure*}
\centering
\includegraphics[width=0.95\linewidth]{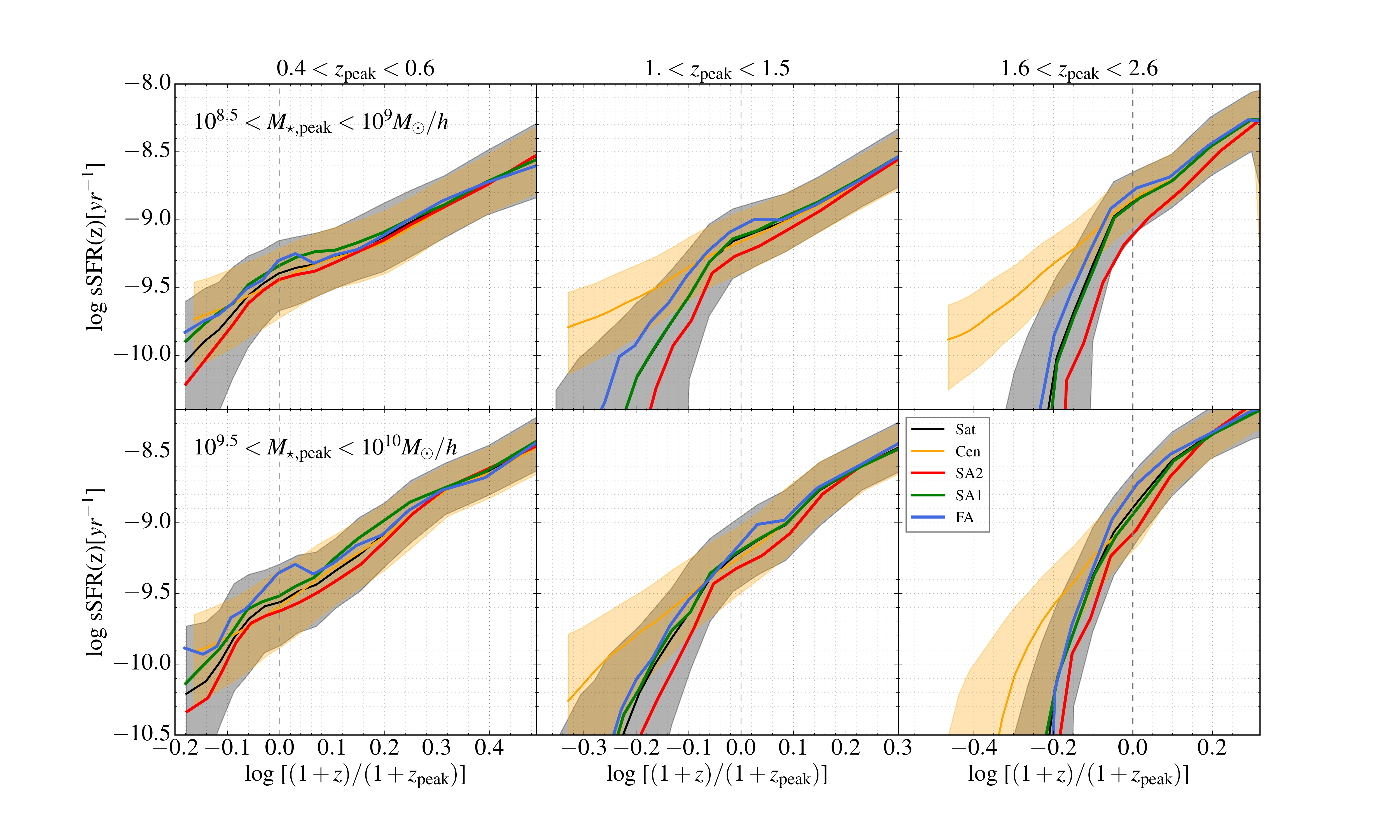}
\caption{\label{fig_sfrz} Similar to figure \ref{fig_mstarz}, but for the sSFR evolution of satellites and centrals from TNG100.}
\end{figure*}

Galaxies grow their mass mainly via two ways, star formation and mergers. In this subsection, we analyze the gas content and star formation of the satellites and how they evolve with time.  In the left panels of Figure \ref{fig_prop_zpeak}, we show the gas to stellar mass ratio and specific star formation rate (sSFR) at accretion time as a function of stellar mass. As before, three $\zpeak$ ranges are adopted. In this plot, the sSFRs of galaxies with a vanishing sSFR value (i.e. sSFR$<10^{-14}yr^{-1}$) are set to be $10^{-14}yr^{-1}$ by hand; galaxies with vanishing gas content are also included when calculating the median values. We can see that galaxies with small stellar mass or at high redshift tend to be more gas rich. This is consistent with previous studies (e.g. \citealt{Genzel2015,Tacconi2018,Calette2018}). The sSFR shows a similar correlation with redshift, but a different correlation with stellar mass. As one can see, sSFR is almost independent of stellar mass at $\log M_{\rm \star,peak}<10$ and then quickly drops as the stellar mass increases. The drop in sSFR at the massive end is caused by the kinetic-mode AGN feedback in the TNG model, which tends to quench star formation \citep{Donnari2019,Terrazas2019,Weinberger2018}.  We also show the results for central galaxies at three snapshots with redshifts close to the considered $\zpeak$ in the right panels. Again, centrals are very similar to these satellites at all stellar mass and redshift bins in consideration. 

We then check whether these properties vary with $\anf$. The results are presented in the right panels of Figure \ref{fig_prop_zpeak}. We investigate the $\anf$ dependence in three $\zpeak$ bins, to make the results free of the effects of the redshift evolution. We find that the satellites in subhalos with the lowest $\anf$ (youngers, blue lines) have largest gas content and the highest sSFR at given $\zpeak$.
This is consistent with results shown in the SMHM relation and mass evolution that the satellites with smaller $\anf$ (younger) grow faster than satellites with larger $\anf$ (older).

Figure \ref{fig_sfrz} shows the evolution of sSFR with time. Since the evolution of gas to stellar mass ratio is similar to sSFR, we only show results for sSFR. Similar to Figure \ref{fig_mstarz}, the results for two $M_{\rm \star,peak}$ ranges and three $\zpeak$ ranges are presented as an example. The results for central galaxies are also presented as a benchmark. First of all, central galaxies exhibit a continuous and steady decline in star formation with redshifts, consistent with previous studies in both simulations and observations (\citealt{VandeVoort2011,Bahe2015,Sales2015}). This is likely due to the continuous decline of gas amount with decreasing redshift (see e.g. Figure \ref{fig_prop_zpeak}). Second, in the central phase, satellites as a whole have almost the same evolutionary history as centrals. This explains why the stellar mass evolution is the same for centrals and satellites in the central phase (see Section \ref{subsec_smhmr}). Third, in the satellite phase ($z<\zpeak$), compared to centrals, the star formation activity in satellites declines much more quickly. This is apparently ascribed to satellite-specific processes, such as strangulation, ram pressure stripping and tidal stripping (\citealt{OstrikerTremaine1975,Peng2015, Yun2019,Bahe2015}). Fourth, in both central and satellite phases, satellites of low $\anf$ (younger) have slightly stronger star formation activities than high $\anf$ (older) ones. Our tests show that, for a given accretion time, low $\anf$ satellites are, on average, more gas rich than high $\anf$ satellites throughout the entire central and satellite phases. It can thus be used to understand why the low $\anf$ satellites grow faster than high $\anf$ ones (see Figure \ref{fig_smhm} and \ref{fig_mstarz}).
Fifth, we observe an apparent enhancement of star formation activity around $\zpeak$ for FA satellites. This enhancement is completely absent for the central galaxies and the SA satellite populations. We also see a similar enhancement in gas to stellar mass ratio. We will come back to this interesting phenomenon later. 

\begin{figure}
\centering
\includegraphics[width=.95\linewidth]{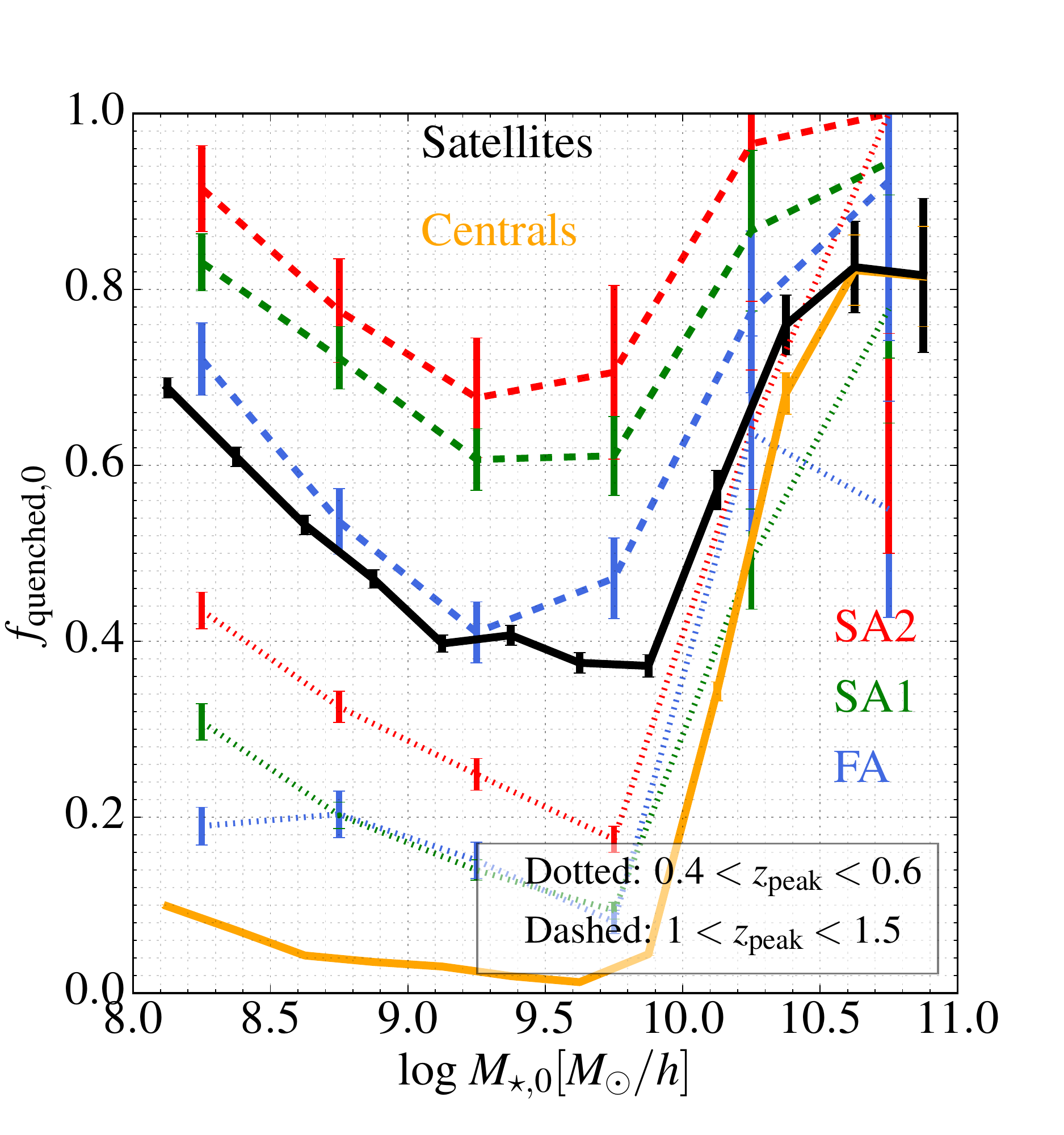}
\caption{\label{fig_fq_z0} 
Quenched fractions at $z=0$ in TNG100. Quenched fractions at $z=0$ as a function of the stellar mass for centrals, satellites as a whole, and the three satellite populations accreted at  $1<\zpeak<1.5$ (dashed) and $0.4<\zpeak<0.6$ (dotted). Satellites in the $1.6<\zpeak<2.6$ bin are almost all quenched at $z=0$, but the results are not plotted for clarity. 
The error bars are Poisson errors.}
\end{figure}

\subsection{Quenched fractions}
\label{subsec_fq}

Finally, we investigate the status of satellites at $z=0$. Since a large fraction of satellites contain no gas particles and have no star formation activity at $z=0$, we show quenched fractions instead of gas content and sSFR. The quenched fraction is the fraction of galaxies with sSFR less than $10^{-11} {yr}^{-1}$ (see e.g. \citealt{Donnari2019} for the discussion of the quenched galaxy definition in TNG100). Figure \ref{fig_fq_z0} shows the quenched fractions as a function of stellar mass. As expected, satellites are more frequently quenched than centrals of the same stellar mass, due to the satellite-specific processes. 
A detailed comparison of satellite quenched fractions with observational data will be useful (Donnari et al. 2020, in preparation). 
 
We also show the quenched fractions for satellites  accreted at two $\zpeak$ ranges with different halo assembly histories separately. As one can see, satellites accreted earlier are more frequently quenched than those accreted later. At a given $\zpeak$, the quenched fraction increases with 
increasing $\anf$, consistent with the results shown above. 

\begin{figure*}
\centering
\includegraphics[width=1.\linewidth]{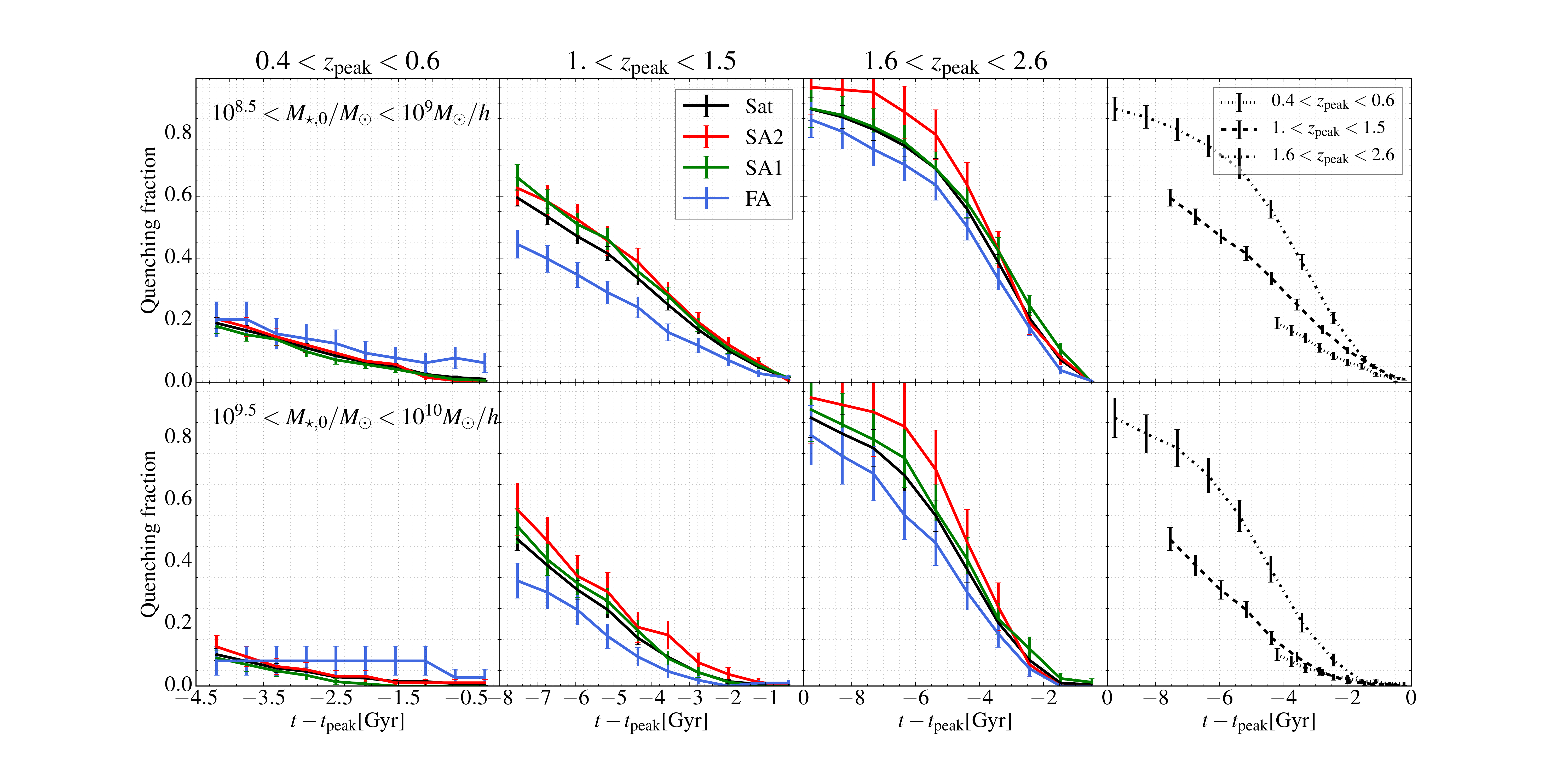}
\caption{\label{fig_fqz} The quenched fractions of satellites as a function of the lookback time relative to the accretion time $t_{\rm peak}$ for satellites of different $\anf$, $\zpeak$ and stellar mass in TNG100. The last column shows the quenching fractions for satellites accreted at three $\zpeak$ bins without distinguishing $\anf$ to highlight the dependence on $\zpeak$.
}
\end{figure*}

To understand more about satellite quenching, we show the quenched fractions of satellite galaxies as a function of look-back time to $t_{\rm peak}$ in Figure \ref{fig_fqz}. As before, we show the results for two stellar mass bins and three $\zpeak$ bins as examples. We can see strong dependences of quenching time scale on stellar mass, accretion time and satellite halo assembly history. More massive satellites are quenched in a longer time scale than less massive satellites for $\log M_{\star,0}<10.0$. This suggests that environmental processes are more efficient for less massive satellites, as is to be expected. Satellites accreted early are quenched more quickly than those accreted later. One possible reason is that the gas environment in host halos at high redshift is more dense than that at low redshift. Therefore, the environmental effects are stronger at high redshift. Consistent with Figure \ref{fig_fq_z0}, the time scale for satellites of high $\anf$ (older) to be quenched is longer than that for satellites of low $\anf$ (younger). There are two possible reasons. First, at accretion time, satellites in lower $\anf$ subhalos are more gas rich than satellites in higher $\anf$ subhalos. Second, in the satellite phase, satellites that are in lower $\anf$ subhalos seem capable of acquiring more cold gas through accretion and even mergers (see Section \ref{subsec_gal_merger}).

\subsection{Galaxy stellar mass accretion and mergers}
\label{subsec_gal_merger}

\begin{figure*}
\centering
\includegraphics[width=1.\linewidth]{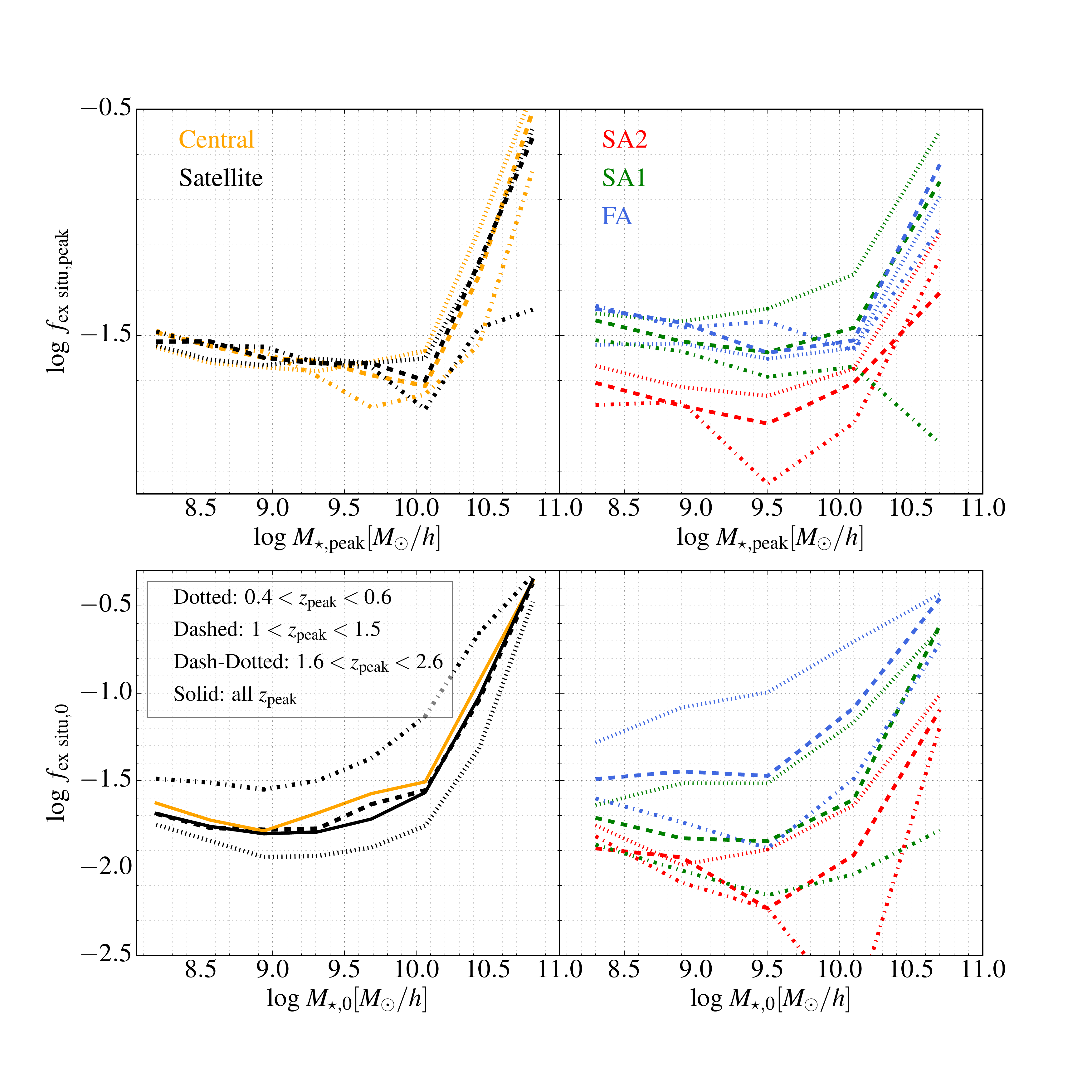}
\caption{\label{fig_facc_mstar} Ex situ stellar mass fractions for TNG100 galaxies. Upper left: the ex situ stellar mass fractions as a function of stellar mass for satellite galaxies at different $\zpeak$ and centrals at corresponding redshifts. Upper right: the ex situ stellar mass fractions as a function of the stellar mass for the three satellite populations at three $\zpeak$ bins. Lower panels: the ex situ stellar mass fractions at $z=0$. The dashed-dotted, dashed, and dotted lines show the results for satellites with different $\zpeak$. }
\end{figure*}

\begin{figure*}
\centering
\includegraphics[width=1.\linewidth]{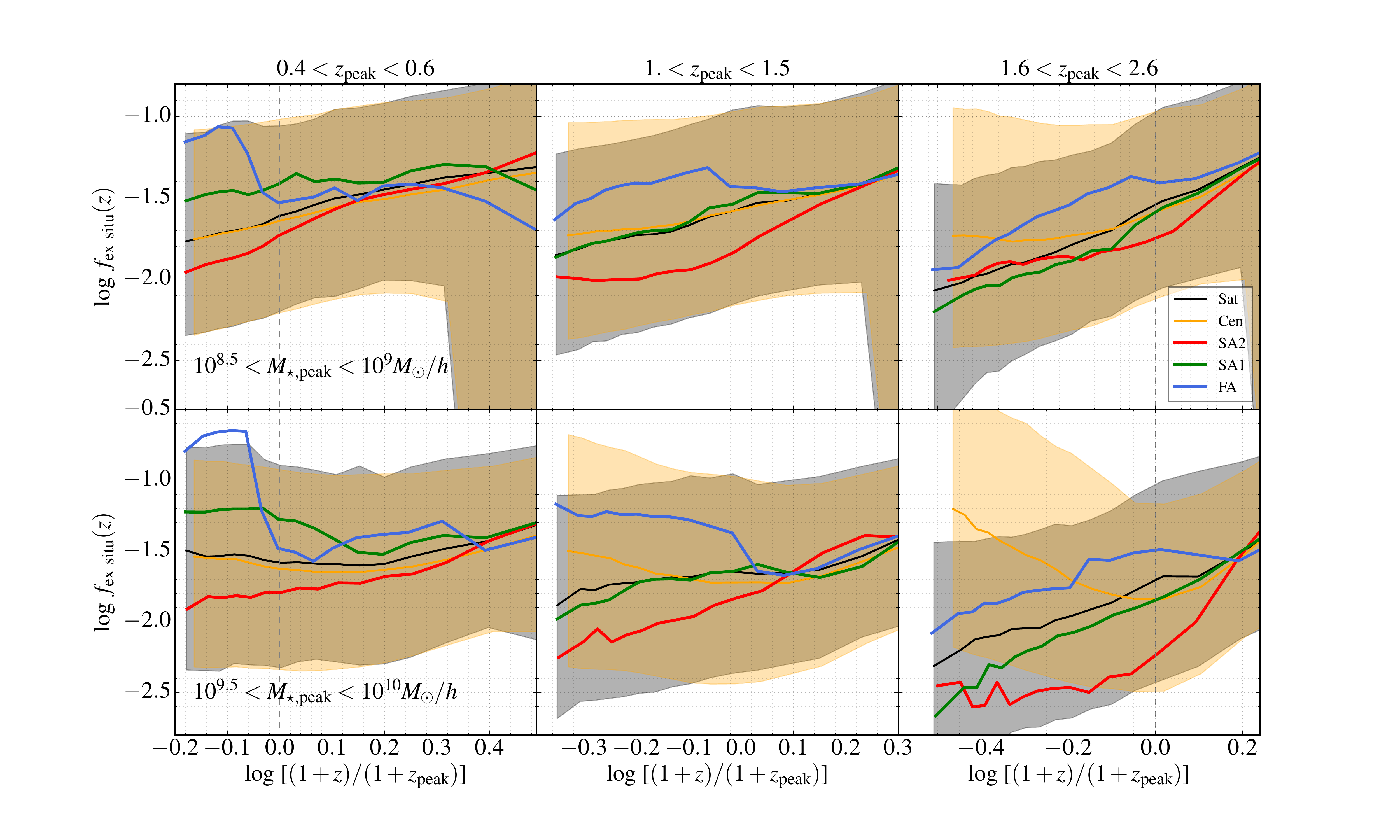}
\caption{\label{fig_faccz} 
Similar to Figure \ref{fig_mstarz}, but for the evolution of the accreted stellar mass fractions (i.e. the ex situ stellar mass fractions) of galaxies in TNG100.}
\end{figure*}

\begin{figure*}
\centering
\includegraphics[width=0.9\linewidth]{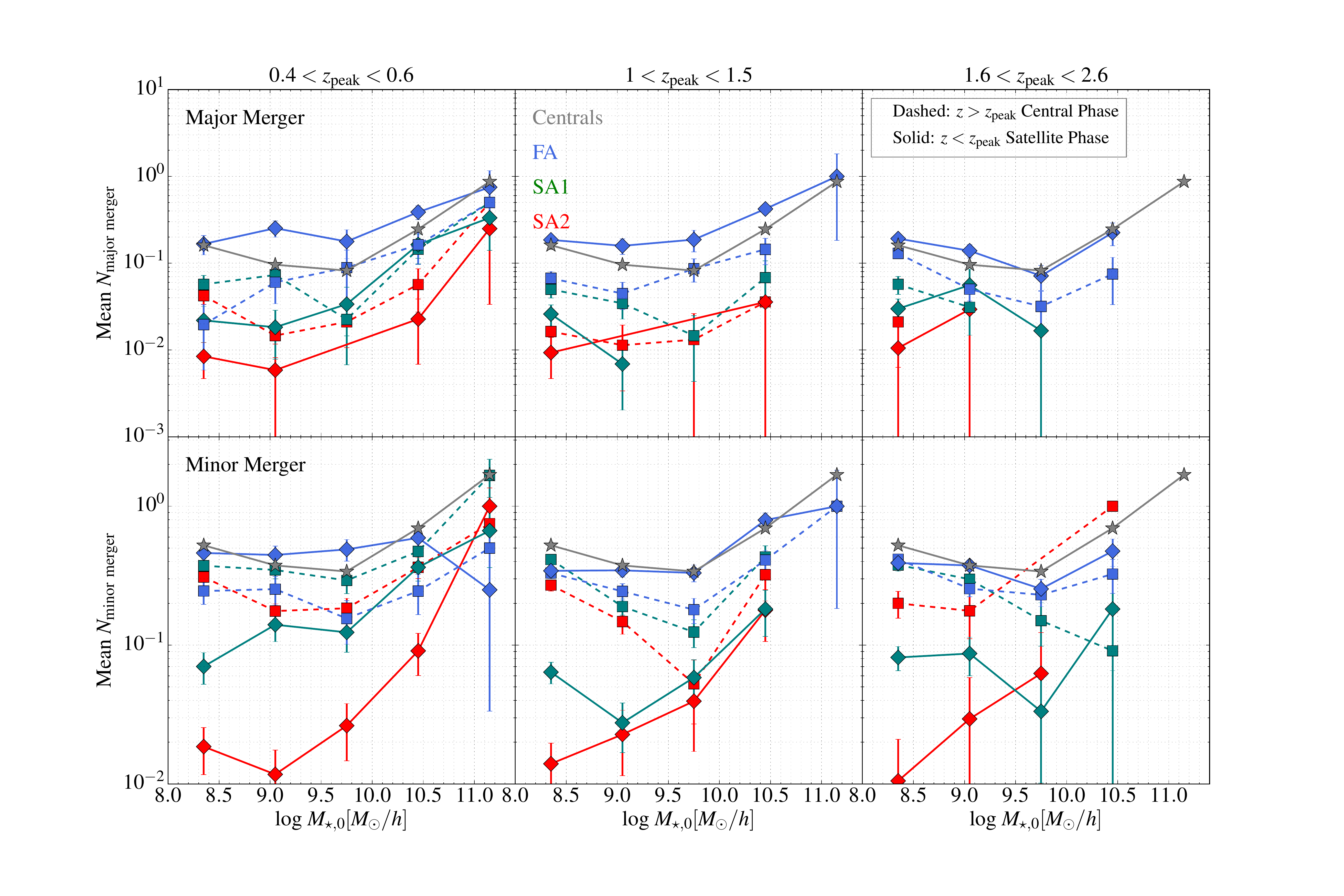}
\caption{\label{fig_nmerger} The mean number of major mergers and minor mergers experienced by the satellites in the three populations of halos (blue, green, and red) accreted at varying $\zpeak$ (from left to right columns) in TNG100. Major mergers are defined as $M_{\rm \star, prog}/M_{\rm \star, 0}>0.2$; Minor merger are defined as $0.05<M_{\rm \star, prog}/M_{\rm \star, 0}<0.2$, where
$M_{\rm \star, 0}$ is the $z=0$ stellar mass of a satellite in consideration, while $M_{\rm \star, prog}$ is the maximum mass of its merging companion. Squares show the results for mergers occurring before $\zpeak$, and diamonds are for mergers experienced after $\zpeak$. The mean merger number of centrals at $z=0$ are plotted as gray stars.}
\end{figure*}

Galaxy mergers play an important role in mass growth and morphological transformation. It is thus interesting to investigate the importance and frequency of galaxy mergers for satellites. We first use $f_{\rm ex\ situ}$ (see Section \ref{subsec_sample} for the definition) to quantify the importance of galaxy mergers.
In Figure \ref{fig_facc_mstar}, we show $f_{\rm ex\ situ}$ at accretion ($f_{\rm ex\ situ, peak}$) and at $z=0$ ($f_{\rm ex\ situ, 0}$) for satellites with different $\zpeak$ and $\anf$. The results for the corresponding centrals are also presented for comparison. On average, $f_{\rm ex\ situ}$ is low, ranging from $0.03$ to $0.3$, indicating that in situ star formation dominates the stellar mass growth for most galaxies with $M_{\star, \rm 0}<10^{10}\msun$ 
(See \citet{LuZhankui2015}, see also
\citealt{Pillepich2018smhmr} for the ex situ stellar mass fractions at larger galaxy masses). One common feature among these centrals and satellites of different $\anf$ at different redshifts is that $f_{\rm ex\ situ}$ is almost constant with stellar mass at $\log M_{\star}<10.0$, and rapidly increases with stellar mass at the massive end.
Such behavior is quite similar to sSFR and quenched fractions shown in Figure \ref{fig_prop_zpeak} and \ref{fig_fq_z0}. One likely reason for the similarity is that, in TNG, galaxy mergers can cause the growth of supermassive black holes and trigger AGN feedback, which can cease star formation in host galaxies. This hints that galaxy mergers play an important (or even dominant) role in triggering AGN feedback and quenching star formation in TNG (see \citealt{Tacchella2019} for an in depth discussion of this in the TNG100 simulation).

As we can see, there is no significant difference in  $f_{\rm ex\ situ, peak}$ between central and satellites. At $z=0$, centrals and satellites also have quite similar $f_{\rm ex\ situ, 0}$ at a given stellar mass, on average. This means that environment has little effect on $f_{\rm ex\ situ}$. However, the dependence of $f_{\rm ex\ situ}$ on $\zpeak$ and $\anf$ appears to evolve with time. At $\zpeak$ time, $f_{\rm ex\ situ, peak}$ is almost independent of $\zpeak$, while at $z=0$, we observe a strong dependence. Moreover, at accretion time, satellites with large $\anf$ tend to have slightly smaller $f_{\rm ex\ situ, peak}$ than their counterparts with small $\anf$. At $z=0$, the difference becomes larger for satellites in a given $\zpeak$ bin. These results imply that some specific processes occur in the satellite phase, as we will see below.

To better understand the results shown above, we present the evolution of $f_{\rm ex\ situ}$ for two $M_{\rm \star,peak}$ bins and three $\zpeak$ bins in Figure \ref{fig_faccz}. Central galaxies exhibit a mass dependent of evolution. For the less massive galaxies, in general, $f_{\rm ex\ situ}$ decreases with time, indicating that in situ star formation is more important at low redshift. 
For massive galaxies selected at low redshift, $f_{\rm ex\ situ}$ almost does not evolve with time, while for massive galaxies selected at high redshift, $f_{\rm ex\ situ}$ first decreases then increases as cosmic time passes. This is apparently the result of the competition between star formation and mergers. These results are broadly consistent with \cite{Rodriguez-Gomez2017}. However, the detailed processes responsible for such evolution is beyond the scope of this paper and we will mainly focus on the behavior of satellites. 

In the central phase ($z>\zpeak$), the difference among centrals, FA, SA1, and SA2 populations are generally small. The significant lower values of SA2 (red lines) in $1.6<\zpeak<2.6$ bin is mainly caused by poor statistics. It is consistent with the results at $\zpeak$ shown in Figure \ref{fig_facc_mstar}. We note that, at $\zpeak\sim0.5$, the results for FA satellites are significantly different from others. This is very likely due to the fact that the FA population at low redshift is relatively small (see e.g. Figure \ref{fig_bimodal}).
In the satellite phase,  satellites deviate from the tracks of centrals and the deviation becomes larger and larger with time. For satellites accreted at low redshift ($\zpeak<1.5$), the difference between centrals and satellites at $z=0$ is small, but for those accreted at $\zpeak>1.6$, the deviation is very large. In fact, we observe a persistent and strong decrease of $f_{\rm ex\ situ}$ for satellites of high $\zpeak$, very different from their central counterparts. In the satellite phase, the stellar mass growth is small and occurs only at the beginning (see Section \ref{subsec_smhmr} and Figure \ref{fig_mstarz}), so the persistent decrease of $f_{\rm ex\ situ}$ cannot be primarily ascribed to the contribution of in situ star formation. One possible reason is that tidal stripping is more efficient for accreted stars since they are usually located in the outer regions of the galaxies. 

In the satellite phase, FA (blue line) and SA (green and red lines) galaxies exhibit very different evolutionary tracks in $f_{\rm ex\ situ}$. We observe a `burst' of $f_{\rm ex\ situ}$ around $\zpeak$ for FA satellites, particularly at low $\zpeak$. This means that a large fraction of the FA population experiences merger events around $\zpeak$. It is usually believed that mergers can enhance the star formation rate \citep{MihosHernquist1996}. Indeed, we also observe an enhancement in star formation activity, as shown in Figure \ref{fig_sfrz}). However, no signal for such merger events is found for the two SA samples. This explains why the difference in $f_{\rm ex\ situ}$ between FA and SA populations becomes larger at $z=0$  (Figure \ref{fig_facc_mstar}).

We can also quantify the importance of galaxy mergers by counting the number of merger events during the history of a galaxy. We define mergers with $M_{\rm \star, prog}/M_{\rm \star, 0}>0.2$ as major mergers and $0.05<M_{\rm \star, prog}/M_{\rm \star, 0}<0.2$ as minor mergers, where $M_{\rm \star, 0}$ is the stellar mass of a galaxy at $z=0$ and $M_{\rm \star, prog}$ is the maximum stellar mass of its merging companion. We note that, the conventional definition for merger types uses the ratio of two stellar masses at the moment when mergers happen. However, one of our main concerns is whether galaxy mergers eventually affect the galaxy at $z=0$, we thus classify major and minor merger by comparing the stellar masses of the merging companion and the final galaxy at $z=0$. 

In Figure \ref{fig_nmerger}, we show the mean number of major and minor mergers as a function of $M_{\rm \star, 0}$ for the three populations separately. In order to know when mergers usually happen, we show the results for mergers happening before and after $\zpeak$ separately. In general, the FA population experiences more major and minor mergers than the two SA populations of the same $M_{\rm \star, 0}$ and $\zpeak$ in both central and satellite phases. For the FA population, both major and minor mergers occur more frequently in the satellite phase ($z<\zpeak$) than in the central phase ($z>\zpeak$). Furthermore, this trend holds in very wide $M_{\rm \star, 0}$ and $\zpeak$ ranges as shown in the figure, while for the two SA populations, the trends are reversed. In addition, it also shows that the difference between the three populations in the central phase is much less than that in the satellite phase, consistent with the results for $f_{\rm ex\ situ}$ (see Figure \ref{fig_faccz}). 

It is particularly interesting to inspect the results in the satellite phase in more detail.  Major mergers for SA satellites with $\log M_{\star, 0}<10.0$ seem rather rare, with mean number of about $0.02$, while for the FA satellites, the mean number of major mergers is about $10$ times higher. Minor mergers occur much more frequently. The mean minor merger number for FA satellites with $\log M_{\star, 0}<10.0$ is about $0.5$, $10$ times higher than that for SA satellites. For comparison, we also show the results for central galaxies at $z=0$. These results suggest that, even in the satellite phase, galaxy mergers occur. In particular for FA satellites, the number of mergers that happen during the satellite phase can be higher/comparable to the total mergers happening for centrals. Note that for central galaxies, we count the merger events in their whole life. It is found that the mean number of major merger for centrals is usually lower than that for FA satellites (even if we count only the mergers in the satellite phase) and higher than that for the SA population. Furthermore, for minor mergers, the mean merger number for centrals is comparable to that for FA satellites and still higher than SA satellites.

\section{Implications for satellite evolution}
\label{sec_dis}

Although hydrodynamical simulations cannot model all the details of the physical processes in galaxy formation, especially those occurring on small spatial scales within the interstellar medium, they still provide valuable information about how galaxies form and evolve, which may be useful for improving other galaxy formation models, such as SHAMs and semi-analytic models (Section \ref{subsec_implication_sham} and \ref{subsec_implication_merger}). In fact, in simulations like TNG100 all the physical mechanisms that are important for the evolution of satellite and central galaxies are emerging phenomena and are not put in by hand: e.g. the hierarchical growth of structure, tidal and ram-pressure stripping, dynamical friction, etc.
Moreover, when analyzing the efficiency of the quenching processes using observational data, we usually adopt some assumptions. Hydrodynamical simulations can also be used to check these assumptions (Section \ref{subsec_implication_quenching}).

\subsection{Implications for SHAM}
\label{subsec_implication_sham}

In the simplest implementation of SHAMs \citep{Kravtsov2004,ValeOstriker2004,Conroy2006}, galaxies and halos (subhalos) are rank ordered by $M_{\rm \star, 0}$ and $M_{\rm halo}$, and matched one by one according to their ranks so that $n_{\rm gal}(>M_{\rm \star, 0})=n_{\rm halo}(>M_{\rm h})$ is satisfied. Here $M_{\rm halo}$ is usually taken as $M_{\rm peak}$ for satellites and $M_{0}$ for centrals. Instead of direct rank ordering, other SHAM implementations \citep{Yang2012,Moster2013,Behroozi2013} parameterize the SMHM relation for centrals at various redshifts.
The parameters in the SMHM relation are constrained using the observed stellar mass function at various redshifts. However, these SHAM methods include several assumptions (see also \citealt{Campbell2018} for a relevant discussion): (i) the rank ordered SHAM (rSHAM) assumes the same $z=0$ SMHM relation for satellites and centrals, and that the SMHM relation is independent of the accretion time of satellites,
(ii) the parameterized SHAM (pSHAM) assumes that satellites share the same $\zpeak$ SMHM relation as centrals, and there is neither gain nor loss of their stellar mass after accretion time, (iii) for both the rank-ordered and parameterized SHAM, at a given halo mass, the stellar mass of galaxies is independent of halo formation history. 

More recently, \cite{Campbell2018} investigated the above halo-mass based SHAM models and found that they generally fail to reproduce the small-scale galaxy clustering signal. The authors tried to relax some of the assumptions in order to save those models, for example, by including orphan satellites, including satellite growth after accretion in pSHAM, and considering the dependence of the SMHM relation on halo formation history. However, none of the above solutions alone can solve the ``small-scale" clustering crisis, indicating a more detailed investigation of these assumptions is needed. Our results can provide a check for these (at least part of) assumptions made in those SHAM methods and help to provide insights for future improvements that can be done from the point of view of the TNG models. 

Our studies (see Figure \ref{fig_smhm}) support the assumptions made in rSHAM that satellites and centrals follow the same $z=0$ SMHM relation when expressed as a function of $\mpeak$ for the satellites, and the relation for satellites is independent of accretion time. Our results further suggest that the scatter in the relation is also the same for centrals and satellites, and that
the assumption that centrals and satellites share the same $\zpeak$ SMHM relation (where $\mpeak$ is adopted as the halo mass for satellites), which is usually adopted in pSHAM. These models usually ignore the mass growth of satellites in the satellite phase, which is apparently in conflict with our results and those by \cite{Engler2020} (based on TNG50, TNG100, and TNG300). However, as shown in Figure \ref{fig_smhm}, the mass growth after accretion does not change the SMHM relation too much when the latter is expressed for the satellites as a function of $\mpeak$. This suggests that considering mass growth after accretion would not improve the model significantly.
Moreover, these pSHAM models usually predict that the $M_{\rm \star, 0}-M_{\rm peak}$ relation for satellites is dependent on accretion time \citep{Campbell2018}, since in these models, the SMHM relation is usually dependent on redshift. In contrast, we find only weak or no dependence on $\zpeak$ in the TNG simulations.

We clearly show that the dependence of the SMHM relation on halo formation time is strong and significant, 
which is ignored by most SHAM models.  In fact, \cite{Matthee2017} and \cite{Artale2018} have already found that the SMHM relation for central galaxies relies on the halo formation time. Our results (Figure \ref{fig_smhm}) demonstrate a clear difference in the SMHM relation for satellites with different halo assembly histories (characterized by $a_{\rm nf}$) at both $z=0$ and $\zpeak$. Since satellites with lower $\anf$ (younger) are usually accreted earlier than satellites with higher $\anf$ (higher), they are expected to reside in the inner regions of host halos at $z=0$. This means that if we use the average SMHM relation to assign satellite mass, it might cause a systematic bias in the ``small-scale" clustering. In addition, caution should be made when one assigns stellar mass to satellites taking into account halo formation history. 
A satellite with low $\anf$ and yet accreted at high $\zpeak$ may have the 
same $\zf$ as an old satellite at low $\zpeak$. Although they have the same $\zf$, 
they have very different SMHM relations (see Figure \ref{fig_smhm}). It thus would be better to adopt $\anf$ rather than $\zf$, since the former takes into account 
$\zpeak$. A quantitative evaluation of how large the effects could be is certainly required.

\subsection{Galaxy mergers for satellites}
\label{subsec_implication_merger}

Galaxy mergers play an important role in galaxy formation and evolution. Major mergers can significantly change the galaxy morphology and may be responsible for the formation of massive ellipticals or bulges (see e.g. \citealt{Torrey2014,NaabBurkert2003,Cox2006}). Moreover, major mergers can trigger starbursts and accelerate the consumption of the cold gas reservoir (e.g. \citealt{MihosHernquist1996,BarnesHernquist1996,Cox2008}). Minor mergers are also important, since they can increase sizes of passive galaxies \citep{Shen2003,Oser2010,Shankar2013}. The galaxy merger scenario is even thought to be able to reproduce the fundamental plane relation \citep{Robertson2006}.

In this paper, we find that satellites can experience frequent merger events. This is the case, however, exclusively for the FA satellites. Even in the satellite phase,  major (minor) merger rate for FA satellites is higher than (comparable to) that for centrals. It means that after $\zpeak$, satellites, exclusively FA satellites, may have even higher (or at least comparable) probability 
to change their morphology or increase their size via mergers than central galaxies of the same stellar mass. Moreover, major mergers likely play a role in consuming the cold gas and quenching star formation in satellite galaxies. More recently, \cite{WangEnci2019} found that the SDSS satellite and central galaxies have similar morphologies as long as halo mass and stellar mass are controlled, while SAM L-galaxies predicted a very different morphology between the two populations. One of the reasons for this discrepancy is that, in SAMs, by construction, mergers between satellites are assumed to be rare compared to the central-satellite merger (see e.g. \citealt{Guo2011}).
Hydrodynamical simulations may provide some clues to improve the treatment of galaxy mergers in SAMs.

As shown in Figure \ref{fig_faccz}, the ex situ stellar mass in the satellite phase (exclusively for FA satellites) usually grows rapidly around $\zpeak$. Most of the ex situ accreted stars can be attributed to merger activities. It is interesting to know why galaxy mergers are boosted at this particular moment. 
FA halos are merging with their host halos at $\zpeak$. 
It is possible that the orbits of the satellites in these FA halos are disrupted by the violent mergers between halos, and then galaxy mergers are enhanced in a short time. However, there is no such signal for the two SA populations, which seems inconsistent with this scenario.  In contrast to FA satellites, mergers for SA satellites in the satellite phase are apparently suppressed compared to central phases (Figure \ref{fig_faccz} and \ref{fig_nmerger}). This means that halo mergers are not necessary to lead to an increase in galaxy mergers. It is likely that whether mergers are enhanced or not depends on the halo formation history or halo inner structure. 
We will discuss this in a subsequent paper.

\subsection{Implications for satellite quenching}
\label{subsec_implication_quenching}

Observationally, attempts have been made to constrain the efficiency of environmental processes by comparing satellites with centrals \citep{Wetzel2013}. The underlying assumption is that satellites at the accretion time share the same properties as centrals. Our results clearly show that centrals and satellites are very similar in the SMHM relation  at $\zpeak$ (Figure \ref{fig_smhm}) and even mass growth history before $\zpeak$ (Figure \ref{fig_mstarz}). They also share similar median gas to stellar mass ratios and specific star formation rates (Figure \ref{fig_prop_zpeak}) and have a similar evolution of sSFR (Figure \ref{fig_sfrz}). Finally, they seem to have quite similar merging histories before $\zpeak$ (Figure \ref{fig_faccz}). 
The simulation thus validates the assumptions adopted in the relevant studies. 

From the evolution of the quenched fraction (Figure \ref{fig_fqz}), we can see that the quenching time scale for satellites is up to a few billion years long. 
For satellites with $8.5<\log M_{\star,0}<9.0$,   
to quench half of them requires about $4$ Gyrs 
for $\zpeak\sim 2$ and about $6.5$ Gyrs for $\zpeak\sim 1.2$. 
For satellites with $9.5<\log M_{\star,0}<10.0$, 
the corresponding time scales increase to 5 Gyrs and 8 Gyrs,
respectively. More than $80\%$ of satellites  
with $\zpeak\sim 0.5$ are still active in forming stars today. 
In particular, we find that, after $2$ Gyrs since accretion 
(i.e. the time of peak mass), only a small fraction of satellites is quenched.
This suggests that these satellites can retain part of their hot gas 
and still acquire cold gas to fuel their star formation even when their 
dark halos stop growing. In practice, it can be expected that these 
time scales for quenching will be shorter and better captured, if the 
time of accretion was chosen to represent the time when 
a satellite crosses the virial radius instead of $\zpeak$ 
(see Donnari et al., in preparation).

\section{Summary}   
\label{sec_sum}

By using the cosmological hydrodynamical simulation TNG100, we studied in detail the formation histories of subhalos and the evolution of satellite galaxies lying therein. We use a scaled formation time, $\anf\equiv (1+\zf)/(1+\zpeak)$, to characterize the formation history of a satellite subhalo before it is accreted by a massive host halo. Here $\zf$ is the half mass formation time and $\zpeak$ is the accretion time. We choose satellites that have peak mass $M_{\rm peak}>10^{10}\msun$ and within host halos of $M_{\rm 0}>10^{11}\msun$.
Our main results can be summarized as follows: 

\begin{itemize}
    \item[(1)] The scaled formation time distribution for subhalos in central phase is bimodal in TNG100. We divide them into three populations, fast accretion (FA, with $\anf<1.3$) and slow accretion (SA), which is sub-divided into SA1 with $1.3<\anf<1.8$ and SA2 with $\anf>1.8$. At a given $\zpeak$, subhalos with larger $\anf$ are older. Satellites accreted at high $\zpeak$ are dominated by the FA population, while satellites accreted at low $\zpeak$ are mainly SA halos. (Figure \ref{fig_bimodal}) 
    
    \item[(2)] When expressed in terms of halo mass at $\zpeak$ for satellites (i.e. $\mpeak$, the central phase peak halo mass), the median SMHM relation at $z=0$ in TNG100 shows no significant difference between centrals and satellites; at accretion time, $\zpeak$, the SMHM relation is also roughly the same between satellites and centrals. Besides, the SMHM relation shows very weak redshift evolution in TNG100. However, the SMHM relations are systematically different for young (lower $\anf$) and old (higher $\anf$) subhalos, particularly at $\zpeak$. (Figure \ref{fig_smhm}) 
    
    \item[(3)] The subhalo formation history has a significant impact on the stellar mass of satellite galaxies.  At fixed halo mass at $\zpeak$, the satellites in subhalos with lower $\anf$ (i.e. residing in younger subhalos) are more massive than those in the subhalos with higher $\anf$ (residing in older subhalos). After accretion, satellites continue to grow in stellar mass, with these in the subhalos with lower $\anf$ growing more comparing to those in the subhalos with higher $\anf$. (see Figure \ref{fig_smhm} and \ref{fig_mstarz}) 
        
    \item[(4)] The subhalo formation history also matters for the gas content and star-formation activities of satellites. At a given stellar mass at $\zpeak$, satellites with lower $\anf$ subhalos (i.e. residing in younger subhalos) are more gas rich and have higher sSFRs than satellites in higher $\anf$ subhalos (Figure \ref{fig_prop_zpeak}). After a short delay since accretion, satellite sSFRs decrease faster than for their central counterparts, indicating the role of host halo environments.
    
    \item[(5)] The quenching time scale depends on stellar mass, accretion time and subhalo formation history. It takes, e.g., on average up to $4$ Gyr for half of those satellites having $\zpeak\sim 2$ and about $6.5$ Gyr for satellites having $\zpeak\sim 1.2$ with $10^{8.5}<M_{\star, \rm 0}<10^{9}\msun$ at $z=0$ to be quenched (Figure \ref{fig_fqz}). Especially for satellites with lower $\anf$ subhalo, the quenching time is longer than satellites with higher $\anf$ subhalo.  This is due to a combination of the higher gas mass in the central phase and the higher probability of mergers in the satellite phase for FA satellites.
        
    \item[(6)] The subhalo formation history has a dramatic impact on the satellite merger history. At a given $\zpeak$, FA satellites experience merger events more frequently than SA satellites in both central and satellite phases. For FA satellites, mergers mainly happen in satellite phase, while for SA satellites, mergers mainly happen in the central phase (Figure \ref{fig_nmerger}). After being accreted by host halos, galaxy mergers for FA satellites are enhanced, while those for SA satellites are suppressed. For FA satellites, we find a burst of galaxy mergers and enhancement in star formation around $\zpeak$, hinting that violent halo mergers might disrupt the orbits of galaxies in FA halos and trigger galaxy mergers. (Figure \ref{fig_facc_mstar} and \ref{fig_faccz}).

\end{itemize}

\section{Acknowledgments}
J. Shi acknowledges the support of the Boya fellow provided by Peking University, helpful discussions with Lizhi Xie on the satellite-specific process and semi-analytic models. J. Shi also acknowledges discussions with Jeremy Tinker on satellite quenching. This work is supported by the National Key R\&D Program of China (grant No. 2018YFA0404503), the National Natural Science Foundation of China (NSFC, Nos.  11733004, 11421303, 11890693, and 11522324), the National Basic Research Program of China (973 Program)(2015CB857002), and the Fundamental Research Funds for the Central Universities. The work is supported by the Supercomputer Center of University of Science and Technology of China and the High-performance Computing Platform of Peking University in China. LCH  was supported by the National Science Foundation of China (11721303, 11991052) and the National Key R\&D Program of China (2016YFA0400702).

\bibliographystyle{mn2e_new}
\bibliography{ref}

\end{document}